\documentclass[prl, amsfonts, twocolumn, nofootinbib,
showpacs]{revtex4}
\usepackage{graphicx, epsfig}

\newcommand{\be}{\begin{equation}}
\newcommand{\ee}{\end{equation}}
\newcommand{\bea}{\begin{eqnarray}}
\newcommand{\eea}{\end{eqnarray}}

\newcommand{\gapp}{\mathrel{\raise.3ex\hbox{$>$}\mkern-14mu
              \lower0.6ex\hbox{$\sim$}}}
\newcommand{\lapp}{\mathrel{\raise.3ex\hbox{$<$}\mkern-14mu
              \lower0.6ex\hbox{$\sim$}}}
\def\bbox{{\,\lower0.9pt\vbox{\hrule \hbox{\vrule height 0.2 cm
\hskip 0.2 cm \vrule  height 0.2 cm}\hrule}\,}}
\begin{document}
\title{Production of black holes and their angular momentum
distribution in models with split fermions}

\author{De-Chang Dai$^1$}
\author{Glenn D. Starkman$^{1,2}$}
\author{Dejan Stojkovic$^1$}
 \affiliation{$^2$ Department of Physics, Case Western Reserve
University, Cleveland, OH~~44106-7079} \affiliation{$2$ Astrophysics
Department, University of Oxford, Oxford, OX1 3RH, UK}

\begin{abstract}
 \widetext
In models with TeV-scale gravity it is expected that mini black
holes will be produced in near-future accelerators.
On the other hand, TeV-scale gravity is plagued with many problems like
fast proton decay, unacceptably large $n-{\bar n}$ oscillations,
flavor changing neutral currents, large mixing between leptons, {\it
etc.}.
Most of these problems can be solved if different fermions are
localized at different points in the extra dimensions. We study the
cross-section for the production of black holes and their angular
momentum distribution in these models with ``split" fermions. We find
that, for
a fixed value of the fundamental mass scale,
the total production cross section is reduced compared with models
where all the fermions are localized at the same point in the extra
dimensions. Fermion splitting also implies that the
bulk component of the black hole angular momentum
must be taken into account in studies of the black hole decay
via Hawking radiation.
\end{abstract}
\pacs{???}
 \maketitle

\section{Introduction}
\indent

There are relatively few robust experimental predictions of quantum
gravity.
One of those few is that particle collisions with center-of-mass
energies
sufficiently greater than the fundamental scale of quantum gravity,
whatever
that may be, should result in the formation of black holes.
Unfortunately, in traditional models of quantum gravity the fundamental
scale of gravity
is the Planck scale, $M_{Pl} \simeq 1.5\times 10^{19}$GeV --
an energy scale well beyond the reach of both particle accelerators and
(because of the rapid decrease in cosmic ray flux with increasing
energy) cosmic ray detectors.
However, over the last decade considerable attention has focused on
models
in which the fundamental quantum gravity energy scale $M_{\star} \gapp
1$TeV \cite{ADD}.  In these models,
the weakness of gravity is due to the existence of extra dimensions
with large volume, $V^{(d)}$,
in fundamental units $\ell_\star^d \equiv (1/M_{\star})^d$.
(Here $d$ is the  number of extra dimensions.)

The characteristic size of the extra dimensions can be as large as
$0.2$mm
(although it is generally much smaller).
This leaves a lot of room for classical description of higher
dimensional objects,
{\it eg.} higher dimensional black holes.
If two incoming particles collide with a center-of-mass (COM) energy
$E_{CM}$ that is greater than $M_*$, and an impact parameter
smaller than the gravitational radius (Schwarzschild radius for a
static or outermost horizon radius for a rotating black hole) corresponding
to $E_{CM}$,
then one expect a black hole of mass $M_{bh} \lapp E_{CM}$ will form.
The exciting possibility is that such black holes might be produced and
studied in near-future accelerator experiments \cite{BHacc}.
For example, the Large Hadron Collider (LHC), due to start operating in
$2007$, will have $E_{CM}\simeq14$TeV. Numerical estimates
\cite{BHacc}
have suggested that it should produce ${\cal O}(10^7)$ black holes per
year
if $M_{\star}\lapp1$TeV.
Black holes might also be produced by ultra-high energy cosmic rays
scattering off nucleons in the Earth's atmosphere \cite{fs}
and be observed using new detectors like the Auger observatory
(however see \cite{ssd}).

Low scale quantum gravity certainly offers a rich opportunity for
higher dimensional phenomenology.  However, if one is concerned
about realistic predictions for actual accelerator processes
({\it eg.} black hole production at the LHC), one must worry about a
host
of unacceptable predictions of low scale quantum gravity models
\cite{problems}.
Low scale gravity is plagued with many problems like fast proton decay,
unacceptably large $n\bar{n}$ oscillations, flavor changing neutral
currents,
large mixing between leptons, {\it etc.}.

One solution to at least some of these problems is to gauge baryon (B)
and lepton (L) number. However, gauging B or L has proven to be
problematic.
If $U(1)_B$ were an unbroken gauge symmetry,
there would be a long range interaction not seen in experiments.
Therefore,  $U(1)_B$ needs to be broken down to a discrete gauge
symmetry.
The leftover discrete symmetry could preserve baryon number modulo some
integer \cite{KraussWilczek}.
To suppress dangerous $n \rightarrow \bar{n}$ (neutron-antineutron)
oscillations
one must forbid both $\Delta B = 1$ and $2$ operators.
The allowed operators of lowest dimension would then be $\Delta B \geq
3$,
and would be of dimension $12$ and higher.
The most common problem in building models with gauged baryon number
is arranging for the cancellation of gauge anomalies \cite{IbanezRoss}.
This requires either an unusual charge assignment to existing particles
or the introduction of new exotic particles.
There are other problems related to the idea of gauge coupling
unification.
The same statements are true for $U(1)_L$ as an unbroken gauge
symmetry.
It is therefore convenient to search for alternative solutions
to the problems of TeV scale gravity.

A widely studied alternative to gauging baryon or lepton number
in order to protect the proton is the so called ``split fermion" model
\cite{ArkaniHamedetal}.
In this model, standard model fields
are confined to a ``thick'' brane -- much thicker than
$M_{\star}^{-1}$.
Quarks and leptons are stuck on different three-dimensional slices
within the thick brane (or on different branes),
separated by much more than $M_{\star}^{-1}$.
This separation causes an exponential suppression
of all direct quantum-gravity couplings (of the type QQQL) between
quarks and leptons,
by resulting in exponentially small wave functions overlaps.
If the spatial separation between the quarks and leptons
is greater by a factor of at least 10 than the widths of their wave
functions,
then the proton decay rate will be safely suppressed.

The splitting of leptons from quarks does not suppress $\Delta B=2$
processes,
like $n-{\bar n}$ oscillations,
which are mediated by operators of the type uddudd.
That requires, for example, further splitting between up-type and
down-type quarks.
Since the experimental limits on $\Delta B=2$ operators are much less
stringent
than those on $\Delta B=1$ operators,
it is enough that the u and d quarks  be separated by
a factor of several times the width of their wave functions.

Splitting between the different quark flavors may have some unexpected
advantages.
Namely, one can explain Yukawa coupling hierarchy by introducing quark
``geography" in extra dimensions \cite{geog}.
Presumably, the Higgs field, H, propagates freely between the sheets
(branes) where quarks, Q, are localized.
Different separations between different quark flavors results in
different wave function overlaps for
operators of the type HQQ involving those flavors, and thus different
effective four-dimensional Yukawa couplings.

The need to suppress both flavor changing neutral currents and
mixing between the neutrino generations requires further
splitting between the different lepton flavors and generations.
Finally, there is a problem of unacceptably large left-handed Majorana
neutrino masses.
This cannot be solved by splitting but requires some other fix.
However, the purpose of this paper is to discuss black hole
phenomenology at the LHC and our discussion is not much affected by the
details of lepton
location in extra dimensions.  We therefore assume that  the black hole
phenomenology will
be unaffected by the particular solution to the neutrino left-handed
Majorana mass problem.

At low energies that do not probe the separation between the fermions,
we must recover the standard ($3+1$)-dimensional results.
These energies are not interesting for black hole production at the
LHC.
However, at energies and momentum transfers high enough to produce
black holes,
one will probe the fermion separations.  One therefore
expects that the ``standard'' ($3+1$)-dimensional black hole
production cross section and angular momentum distribution will be
significantly modified.

After a black hole is formed ({\it eg.} at the LHC), it decays by
emitting Hawking radiation with temperature $T  \sim 1/r_h$,
where $r_h$ is the horizon radius of the black hole.
Thermal Hawking radiation consists of two parts:
(1) particles propagating along the brane, and (2) bulk radiation.
The bulk radiation includes bulk gravitons.
The bulk radiation is usually neglected.
The justification is as follows.
The wavelength of emitted radiation is larger than the size of the
black hole,
so the black hole behaves as a point radiator, radiating mostly in
s-wave.
Thus, the radiation for each particle mode will be equally probable in
every direction (brane or bulk).
For each particle that can propagate in the bulk there is a whole tower
of bulk Kaluza-Klein excitations,
but, since each excitation is only weakly coupled (due to small wave
function overlap) to the small black hole,
the whole tower counts only as one particle.
Since the total number of species that are living on the brane is
quite large ( $\sim 60$)
while there is only one graviton, radiation along the brane should be
dominant (see {\it eg.} \cite{EHM}).

This reasoning works very well if the black hole is not rotating.
Rotation can significantly modify this conclusion.
For high energy scattering of two particles with a non-zero impact
parameter,
the formation of a rotating black hole is much more probable
than the formation of a non-rotating black hole.
One expects that mainly highly rotating mini black holes will be formed
in such scattering.

The number of graviton  degrees of freedom in $(n+1)$-dimensional
space-time is $N=(n+1)(n-2)/2$, which is just the number of possible
polarizations of a spin 2 particle in an $n$-dimensional space.
For example, for $n+1= 10$ we have $N=35$.
If a black hole is non-rotating, we expect emission of particles with
non-zero spin ({\it eg.} gravitons)
to be suppressed with respect to emission of scalar quanta as happens
in $(3+1)$-dimensional space-time
\cite{Page:76} (see also Section~10.5 \cite{FrNo} and references
therein).
However, rotating black holes exhibit the phenomenon of super-radiance.
Due to the existence of an ergosphere (a region between the infinite
redshift surface and the event horizon),
some of the modes of radiation get amplified, taking away the
rotational energy of the black hole.
Super-radiance is strongly spin-dependent, and emission of higher spin
particles is strongly favored.
For an extremal rotating black hole, the emission of gravitons is a
dominant effect.
For example, $(3+1)$-dimensional numerical calculations done by Page
\cite{Page:76} (see also \cite{FrNo})
show that the probability of emission of a graviton by an extremal
rotating black  hole
is about 100 times higher than the probability of emission of a photon
or neutrino.
In \cite{Kerr5D}, it was shown that super-radiance also exists in
higher dimensional space-times.
Mini black holes created in high energy scattering are expected to have
high angular momentum.
We therefore expect a much higher proportion of the initial black hole
mass to be dissipated as bulk gravitons.

It is known that in the highly non-linear, time-dependent and violent
process of black hole creation,
much of the initial center of mass energy is lost to gravitational
radiation.
This is up to 30\% in $3+1$ dimensions, and may be larger in higher
dimensions,
due to the larger number of gravitational degrees of freedom.
Since gravitons are not bound to the  brane, most of them would be
radiated in the bulk.
Bulk graviton radiation may well dominate over the radiation of other
particles in the brane,
at least in the first stages of black hole evaporation.

For an observer located on the brane,
the first signature of bulk graviton emission is missing energy in the
detector.
Also, as a result of the bulk graviton emission, the black hole will
generically
recoil in bulk directions.
In the canonical context of a thin brane, this recoil can move the
black hole off the brane,
unless it is bound to the brane by some other force.
(In some Randall-Sundrum type models this bulk recoil is forbidden by a
$Z_2$ symmetry.)
After the black hole leaves the brane, it cannot emit brane-confined
standard model particles anymore.
Black hole radiation would be abruptly terminated for an observer
located on the brane.
The probability for something like this to happen depends on many
factors (black hole mass, brane tension...) and was studied in
\cite{recoil,flux}.

In brane-world models in which standard model fermions are localized on
a thin brane
with a delta function wave function, a black hole that is formed in
collisions of
the standard model particles will have only a brane component of its
angular momentum.
Bulk graviton emission, as discussed above, if it is not  s-wave, can
also
give the black hole a non-zero bulk component of angular momentum.

If, however, the localization wave function  has a spread of order
$\Delta \simeq M_\star^{-1}$,
then black holes made in collisions of relativistic particles with
$E_{CM}\simeq M_\star$
will have bulk angular momenta of order $\Delta E_{CM} \simeq 1$.

In the split fermion model the initial bulk component of the angular
momentum of the initial particles can be quite large,
since the quarks are localized at different points in the extra
dimensions.
Thus the black hole made in such a collision would be expected to have
a non-zero bulk angular momentum.
Bulk emission by such black holes will be significant.

In this paper we consider the consequences of this possibility that
the black holes formed in high energy collisions will have appreciable
bulk and brane angular momentum.
We look at both the production cross-section and the angular momentum
distribution.

\section{An Illustrative Model}
\label{illustrative}

It is only in the context of a definite model that we can perform
explicit calculations.
As we shall see there is enormous freedom in  the properties of models.
We shall therefore begin by  defining a particular model,
proceed to calculate within that model,
and then attempt to distill from the  results those features that are
generic.

We start with the quarks.
There is a freedom of exactly where in the extra dimensions
we should localize the various quarks.
For the sake of definiteness, we adopt the phenomenologically motivated
scheme from \cite{geog}
where the quark separations are set to reproduce the hierarchies
present
in the Yukawa couplings of the standard model.

\begin{figure}[h]
    \centering{
    \includegraphics[width=3.0in]{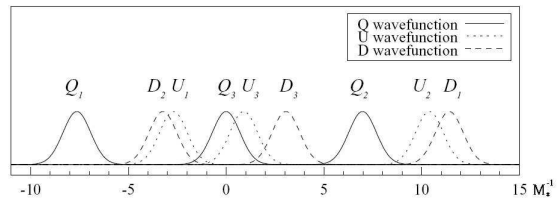} }
    \caption{Fermion distribution in extra dimensions in the
illustrative model of \cite{geog}.}
    \label{fig:fermionlocation}
\end{figure}

Left and right-handed quarks of each flavor are in different locations.
These are given
in fundamental units in table 1, and and are depicted in figure
\ref{fig:fermionlocation}.
\begin{table}
\caption{The positions (in fundamental units) of the quark fields in
the illustrative model of \cite{geog}.}.
\label{tab:positions}
\begin{tabular}{l|r}
\hline
Quark type & Position ($M_\star^{-1}$) \\
\hline
$Q_1=(u,d)_L$  & -7.6067  \\
$Q_2=(c,s)_L$  & 6.9522  \\
$Q_3=(t,b)_L$  & 0.0  \\
$U_1=u_R$  & -2.7357  \\
$U_2=c_R$  & 10.4362 \\
$U_3= t_R$  & 0.9012  \\
$D_1= d_R$  & 11.3682  \\
$D_2= s_R$  & -3.2250 \\
$D_3= b_R$  & 3.0511 \\
\hline
\end{tabular}
\end{table}

We also have some freedom in choosing the extra-dimensional profile of
the quark wave functions.
We will consider two cases:
(1) each quark is sharply localized at a single point in the extra
dimensions
(i.e. Dirac delta function profile)
and
(2) the quarks have bulk wave functions with a Gaussian profile.
In case (2), we take the width  of the Gaussian profile to be
$\Delta=M_\star^{-1}$.

Since Higgs and gauge bosons (gluons, W's, Z's and photons) must
interact with all the quarks (and leptons) they can not be localized at
one point
but have to propagate between the sheets where the fermions are
localized. We take the wave functions of these particles to be uniform
across the full brane.

Finally, gravitons are not confined to the mini-bulk defined by
the standard model thick brane (which may contain several sub-branes
where different fermions are localized), but can propagate
everywhere in the full bulk.

\section{Black hole production cross section}
\indent

Consider two  particles colliding
with a center of mass energy $E_{CM}$.  They will also have an angular
momentum $J$
in their center of mass frame.
If the impact parameter between the two colliding particles is smaller
than the
diameter of the horizon of an $(n+1)$-dimensional black hole of  mass
$M=E_{CM}$
and angular momentum J,
\be
b < 2 r_h(n,M,J) ,
\ee
then a black hole will form \cite{BHacc}.
We can say that the incoming particles interacted via non-linear
quantum gravity interactions.
Since the incoming particles in question are highly relativistic
(therefore there is no gravitational focusing),
the cross section for this process is precisely equal to the
interaction area $\pi (2r_h)^2$.


In our case, where the incident particles are separated in both the
ordinary and the extra dimensions,
they will have a non-zero impact parameter $b_e$ in the bulk
directions.
No black hole will therefore be formed if their minimal
four-dimensional separation (along the brane) exceeds
$\sqrt{4 r_h^{2}(n,E_{CM},J)-b_e^{2}}$. The black hole production
cross section will therefore be
\be
\sigma_{production}(E_{CM},J) = \pi (4 r_h^{2}(n,E_{CM},J)-b_e^{2})  .
\ee
This cross section is smaller than  the usually quoted $\pi
4r_s^2(n,E_{CM})$ ,
both because of the effect of the extra-dimensional separation, $b_e$,
and because the horizon radius, $r_h$,
is smaller than the Schwarzschild radius, $r_s$, whenever $J\neq0$.

\subsection{The impact parameter}
\label{impactparameter}
\indent

The total black hole production cross section depends on the maximal
impact parameter between two particles that can yield a black hole of a
certain mass --
twice the horizon radius of the black hole with the combined
center-of-mass energy and angular momentum of the two particles.
We must therefore determine this horizon radius.

In Boyer-Lindquist coordinates,
the metric for an $(n+1)$-dimensional black hole
with angular momentum parallel to the z-axis is:
\begin{eqnarray}
ds^2 &=&\left( 1-\frac{\mu r^{4-n}}{\Sigma (r,\theta)}\right)dt^{2}
\nonumber \\
&-& \sin^{2}\theta \left(r^{2}+a^{2}\left(
    + \sin^{2}\theta \frac{\mu r^{4-n}}{\Sigma (r,\theta
)}\right)\right)
    d\phi ^{2} \nonumber \\
&+& 2 a \sin^{2}\theta \frac{\mu r^{4-n}}{\Sigma (r,\theta )} dt d\phi
\nonumber \\
&-& \frac{\Sigma}{\Delta} dr^{2}-\Sigma d\theta ^{2} - r^{2}cos^2\theta
d^{n-3}\Omega
\end{eqnarray}
where
\begin{eqnarray}
&&\Sigma =r^2+a^{2}cos^{2}\theta \nonumber \\
&& \Delta=r^2+a^{2}-\mu r^{4-n} .
\end{eqnarray}
\begin{equation}
M=\frac{(n-1) A_{n-1}}{16\pi G_n}\mu
\end{equation}
is the mass of the black hole, and
\begin{equation}
J=\frac{2Ma}{n-1}
\end{equation}
is its angular momentum.

Here
\be
A_{n-1} = \frac{2 \pi^{(n-1)/2}}{\Gamma((n-1)/2)}
\ee
is the hyper-surface area of a $(n-1)$-dimensional unit sphere.

The higher dimensional gravitational constant $G_n$ is defined as
\begin {equation}
G_{n}=\frac{\pi^{n-4}}{4 M_{\star}^{n-1}}
\end{equation}

The horizon occurs when $\Delta =0$. That is at a radius given
implicitly by
\begin{equation}
\label{horizon}
r^{(n)}_h=\left[\frac{\mu}{1+(a/r^{(n)}_h)^{2}}\right]^{\frac{1}{n-2}}
=\frac{r^{(n)}_s}{\left[1+(a/r^{(n)}_h)^{2}\right]^{\frac{1}{n-2}}}.
\end{equation}
Here
\be
r^{(n)}_s=\mu ^{1/(n-2)}.
\ee
is the Schwarzschild radius of an $(n+1)$-dimensional black hole.
That is
\begin{equation}
r^{(n)}_s(us,n,M_{pl})=k(n)M_{pl}^{-1}[\sqrt{us}/M_{pl}]^{1/(n-2)}
\end{equation}

\begin{equation}
k(n)\equiv\left[2^{n-3}{\pi}^{(n-6)/2}\frac{\Gamma[n/2]}{n-1}\right]^{1/(n-2)}
\end{equation}

If two highly relativistic particles collide with center of mass energy
$E_CM$, and total impact parameter $b$, then their angular momentum
in the center of mass frame before the collision is $J_{in}=b E_{CM}
/2$.
Suppose that the black hole is formed initially retaining all this
energy and angular momentum.
Then the mass and angular momentum of the black hole will be
$M=M_{in}=\sqrt{s'}$ and  $J=J_{in}$.
A black hole will form if:
\begin{equation}
b < 2r^{(n)}_h (M_{in},J_{in}) \, .
\end{equation}
Therefore the maximum impact parameter satisfies
\begin{equation}
b_{max}  =  2r^{(n)}_h (E_{CM}, b_{max}E_{CM}/2) \, .
\label{bmax1}
\end{equation}
We see that $b_{max}$ is a function of both $E_{CM}$ and the number of
dimensions, $n$ \cite{ida}.

Using equation (\ref{horizon}), we can rewrite condition (\ref{bmax1})
\be
0 = \left(\frac{b_{max}}{2}\right)^2 +
\left(\frac{n-1}{4}{b_{max}}\right)^2 -
\mu \left(\frac{b_{max}}{2}\right)^{-d+1} .
\ee
We thus obtain for the maximum value of the impact parameter
\begin{equation}
b_{max}(E_{CM};n)=2\frac{r^{(n)}_s(E_{CM})}{\left[1+\left(\frac{n-1}{2}\right)^{2}\right]^{1
\over {n-2}}} \,
.
\label{bmax2}
\end{equation}

This result was obtained in \cite{ida}.

\subsection{Cross section when quarks are not separated in bulk
directions}
\indent

If the quarks are not separated in the bulk directions (i.e. $b_e=0$),
but rather all the standard model particles are localized at a single
point in the bulk,
then the geometric cross section for black hole production is $\pi (4
r_h^2(n,E_{CM},J))$.
The presence of $d$ extra dimensions changes only the value of the
horizon radius,
as per equation (\ref{bmax2}).

At the LHC, each proton will have $E=7$TeV in the COM frame.
Therefore, the total proton-proton center of mass energy
will be $\sqrt{s} = 14$TeV.
If two partons have energy $vE$ and $\frac{uE}{v}$,
much greater than their respective masses, then the parton-parton
collision will have
\begin{equation}
s'=|p_{i}+p_{j}|^{2}=|v(E,E)+\frac{u}{v} (E,-E)|^{2}=4uE^{2}=us \, .
\end{equation}
The center of mass energy for these two partons will be $\sqrt{us}$,
as will be the 4-momentum transfer $Q^{2}$.
The largest impact parameter between the two particles that can form a
black hole with this mass will be $2r_h$.

The total proton-proton cross section for black hole production is
therefore
\begin{eqnarray} \label{partcross}
&&\sigma^{pp\to BH}(s;n,M_\star)  =
\int_{M_\star^2/s}^{1}\!\!\!\!du \int_{u}^{1}\frac{dv}{v}
    \pi \left[b_{max}(\sqrt{us})\right]^2 \nonumber \\
    &&\quad\quad \times \sum_{ij}
f_{i}(v,Q=\sqrt{us})f_{j}(u/v,Q=\sqrt{us}) \, .
\end{eqnarray}

Here $f_i(v,Q)$ is the i-th parton distribution function.
Loosely this is the expected number of partons of type i and momentum
$vQ$
to be found in the proton in a collision at momentum transfer $Q$.
This result was also obtained in \cite{ida}.

\subsection{Cross section when quarks are separated}
\indent

Next, we consider the case when two partons can be located in different
points
of the extra dimensions.
Denote the wave functions of the particles in the extra dimensions as
$f^{(e)}_{i}({\vec x_e})$  and $f^{(e)}_{j}({\vec x_e})$
(normalized such that $\int \vert f^{(e)}_{i}(x_e)\vert^2 d^{n-3}x_e =
1$).
The probability that the particles' separation in the extra dimensions
will be $b_e$
is therefore

\begin{equation}
P^{ij}_e(b_e)= \int  b_e^{n-4}d^{n-4}\Omega_{b_e}
    \int d^n x_e \left\vert
f_{ei}(\vec{x_e})f_{ej}(\vec{x_e}+\vec{b_e}) \right\vert^2 .
\end{equation}

The $(n-3)$-dimensional integral $\int d^{n-3} x_e$ is over all
possible
positions ${\vec x_e}$
in the extra dimension; the angular integral , $\int
d^{n-4}\Omega_{b_e}$ ,
is over all extra-dimensional orientations of ${\vec b_e}$.
$P_e$ can be evaluated analytically  in simple cases, such as for
Gaussian wave functions.

For a given impact parameter $b$ in the ordinary dimensions,
$b_e$ can take all possible values from zero to
$\sqrt{b_{max}(s'=us;n)^2-b^{2}}$.
The contribution to the total cross section from the impact parameter
interval
$[b,b+db]$, must therefore be weighted  by
\begin{equation}
w^{ij}_e(b) = \int_{0}^{\sqrt{b_{max}(\sqrt{us})^{2}-b^{2}}}P^{ij}_e(b_e)db_e \,.
\end{equation}

Applying this to equation (\ref{partcross} ) we obtain
\begin{eqnarray} \label{aa}
\sigma^{pp}(s,n,M_\star)  &=
\sum_{ij} \int_{M_\star^2/ s}^{1}du \int_{u}^{1}\frac{dv}{v}
\int_0^{b_{max}(us;n)} w^{ij}_e(b) 2\pi b db
    \nonumber \\
    & \times  f_{i}(v,Q)f_{j}(u/v,Q)
    \, .
\end{eqnarray}

\subsection{The Differential Cross-section, $\frac{d \sigma}{dJ}$}
\indent

The symmetry group of rotations in $n$ spatial dimensions, $SO(n)$ has
$\left[n/2\right]$ (the integer part of $n/2$) Casimir operators.
This means that there are $\left[n/2\right]$ independent planes of
rotations with
the same number of parameters of rotation. In the usual
($3+1$)-dimensional case,
$n=3$, so there is only one parameter of rotation.

In general, angular momentum is defined as:

\begin{equation}
\label{high angu}
J^{ij}=\int d^{N}x (x^{i}T^{j0}-x^{j}T^{i0})
\end{equation}
where $T^{ij}$ is the momentum density.


Thus, in general, there is a single plane of rotation located within
the brane directions
and can be one or more planes of rotation extending into the extra
dimensions.
However, in the case at hand, there are only two particles involved in
the collision, with linear momentum along the brane.
They move towards each other along two lines that, assuming a non-zero
impact parameter, define a single plane.
Therefore, assuming our black hole initially carries only the energy
and angular momentum of these progenitors,
we can always redefine our coordinate system so that the black hole has
only one plane of rotation
(with a single parameter of rotation).
Of course, if $b_e \neq 0$, this plane of rotation will be at an angle
with respect to the brane.
There will therefore be a non-zero bulk component of the angular
momentum.

Consider two relativistic particles with center-of-mass energy
$s'=\sqrt{us}$.
In the center of mass frame, each has energy $s'/2$, and momentum $\pm
s'/2$.
The angular momentum of the system about its center of mass is
\begin{equation}
\label{E7}
J = 2 \frac{s'}{2} \frac{b}{2} = \frac{\sqrt{us}b}{2} .
\end{equation}
Since
\begin{equation}
\label{E13}
\left(\frac{d\sigma}{db}\right)_{\!\!s'} = 2\pi b,
\end{equation}
therefore
\begin{equation}
\label{E14}
\left(\frac{d\sigma}{dJ}\right)_{\!\!s'} = \frac{4\pi b}{\sqrt{us}}  =
\frac{8\pi J}{us}
\end{equation}
Eq. (\ref{E14}) is valid for any two partons.
For the total cross section, we need to sum over all partons and
integrate over all parton momenta.

Combining  Eq. (\ref{aa}) and Eq. (\ref{E14}),
the cross section change with the brane component of angular  momentum
becomes
\begin{eqnarray}
\label{E9}
\left(\frac{d\sigma ^{pp}}{dJ}\right)_{\!\! s} &=&
\sum_{ij}\int_{\max(\frac{M_\star^{2}}{s},\frac{4J^2}{b^{2}s})}^{1}
du\int^{1}_{u}\frac{dv}{v} \frac{8\pi
J}{us}
w^{ij}_e\left(b=\frac{2J}{\sqrt{us}}\right)\nonumber \\
&\times&f_{i}(v,Q=\sqrt{us})f_{j}(u/v,Q=\sqrt{us})
\end{eqnarray}

Similarly, the bulk component of the angular momentum is
\begin{equation}
\label{E8}
J_e=\frac{\sqrt{us}b_e}{2} \, .
\end{equation}
Therefore,
\begin{eqnarray}
\label{E10}
\left(\frac{d\sigma ^{pp}}{dJ_e}\right)_{\!\! s} \!\! &=&\!\!
\sum_{ij}\int_{\frac{M_\star^{2}}{s}}^{1} du\int^{1}_{u}\frac{dv}{v}
\left[ P^{ij}_e\left(b_e\right) \pi \left(b_{max}(\sqrt{us})^2 -
b_e^2\right)
    \right]
\nonumber \\
&\times&f_{i}(v,Q=\sqrt{us})f_{j}(u/v,Q=\sqrt{us}) ,
\end{eqnarray}
where $b_e=2J_e/\sqrt{us}$.

%
%
\section{Results}
\subsection{Cross-section decreases with inter-quark spacing}

As two partons are separated in the extra dimensions,
their maximum separation along the brane directions becomes smaller,
as does the associated black hole production cross section.
Thus, the black hole  production cross-section is reduced in
split-brane models compared to the case where all the quarks and gluons
are localized on
the same thin brane.

\begin{figure}[h!]
    \centering{
    \includegraphics[width=3.5in]{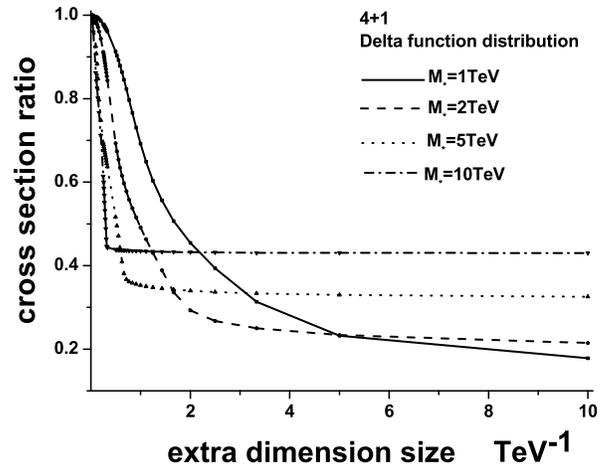}
    }
    \caption{Ratio of black hole production cross-section
    for the $(4+1)$-dimensional split fermion model to the black
    hole production cross-section
    for a $(4+1)$-dimensional model with all the standard model
particles localized at a single point,
    ($\sigma _{split}/\sigma_{no split}$) as a function of the size of the extra dimension.
    The fermion wave functions in both cases are taken to be
delta-functions.
    In the split fermion model, the gluons are permitted to move freely
in the thick brane.
    }
    \label{cro4+1delta}
\end{figure}

\indent

As previously mentioned,
the extra-dimensional profile of quark wave functions is not well
known.
We consider two cases: delta-function profiles, and Gaussian profiles.

We first consider the case where the quarks are sharply localized at
some point in the extra-dimensional space, with Dirac delta function
wave-functions.
Gluons can freely propagate between the sheets where quarks are
localized,
and are taken to have flat (uniform) profiles. The relative location of the quarks
is initially set as in our illustrative split-fermion model (section ``An illustrative model").
We want to study how the cross section decreases  as the separation between the quarks
(which we call the size of the extra dimension) increases. 
So, we change the size of the extra dimension
by rescaling the distance between the quarks, while keeping the fundamental energy scale $M_\star$ fixed. In Fig. \ref{cro4+1delta}, we plot,
as a function of the size (in units of $(1{\rm TeV})^{-1}$) of a single
extra dimension,
the ratio of the black hole production cross section in the split-fermion model
to the cross-section  in a model with all standard model particles
localized at a single point.
We do this for a variety of fundamental energy scales from $1$TeV to
$10$TeV.

If the size of the extra dimension is smaller than the size of the
black hole produced,
then the split fermion model is indistinguishable from the un-split
model.
Therefore the ratio of the cross section in these two cases approaches
unity.
As the size of the extra dimension  increases,  and the separation
between fermions increases in the  split-fermion model,
the maximum $(3+1)$-dimensional impact parameter that results in black
hole creation decreases.
(It is $\sqrt{4r_h^{2}-b_e^{2}}$ as opposed to $r_h$ in the unsplit
model.)
Therefore, the cross section ratio decreases.
The production of smaller mass black holes is affected more than the
production of large black holes.

The higher $M_\star$, the more rapidly the split fermion black hole
production cross section
decreases with the increasing size of the extra dimension.
This decline ceases when the size of the extra dimension exceeds the
size of the black hole.
Beyond that point, fermions that are not at the same location in the
extra dimension
will be too far to interact to form a black hole.
Gluons will still contribute but their interaction is also suppressed
by the size of the extra dimension.
For $(4+1)$ dimensions, the suppression factor is
the ratio of the  black hole horizon $r_h$ to the size $L$ of the extra
dimension.
Interactions among fermions that are co-located in the extra dimensions
will clearly make the dominant contribution, {\it eg.} uu and dd type
of interaction.
The cross section ratio therefore becomes a fixed number,
which is however different for different values of $M_\star$, because
of energy dependent PDF's. The reason
is that as $M_\star$ increases, gluon contribution decreases in the non-split
case, and since the non-split cross section is in the denominator, the
cross section ratio increases.

\begin{figure}[h!]
    \centering{
    \includegraphics[width=3.5in]{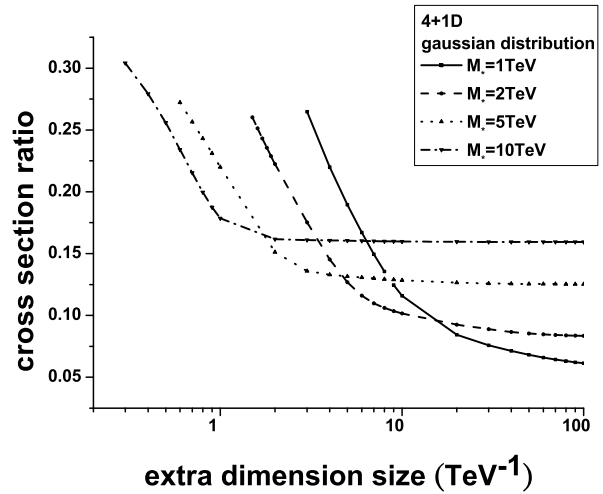} }
    \caption{As in figure \ref{cro4+1delta}, we show the ratio of black
hole production cross-section
    for the $(4+1)$-dimensional split fermion model
    to the black hole production cross-section
    for a $(4+1)$-dimensional model with all the standard model
particles localized at a single point,
    ($\sigma _{split}/\sigma_{no split}$) as a function of the size of the extra dimension.
    Unlike in figure \ref{cro4+1delta}, the fermion wave functions are
taken to be Gaussian.
    }

    \label{cro4+1}
\end{figure}

We next consider the case where the fermion  wave-functions in the
extra dimension are Gaussian.
The gluons can move freely in the extra dimension
between the sheets where quarks are localized,
and are taken to have flat (uniform) profiles..
As before, so long as the size of the extra dimension is very small,
the ratio of the black hole production cross-section in this split
brane model to the production cross
section in a single brane model is very close unity.
At an intermediate size of the extra dimension,
the cross section in the split fermion model is again suppressed due to
the reduced interaction range.
As expected, the decay is slower for the Gaussian wave function than
for the delta function.
For larger distances between the fermions the ratio decays again with
$r_h/L$.
These features can be seen in Fig. \ref{cro4+1}.

\begin{figure}[h!]
    \centering{
    \includegraphics[width=3.5in]{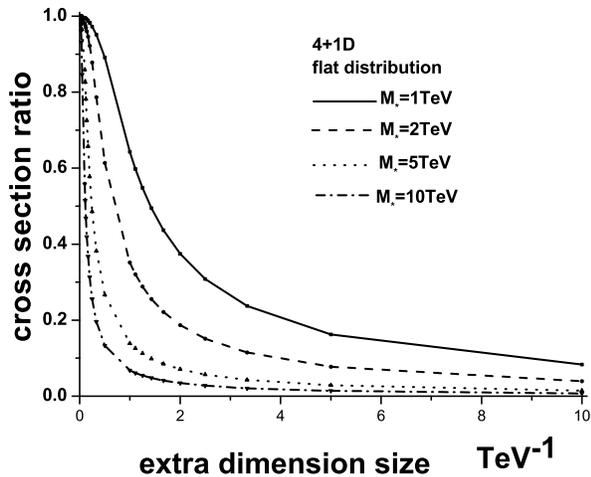}
}
    \caption{This figure shows the ratio of black hole
production cross-section for the $(4+1)$-dimensional thick brane model
where all the standard model particles have uniform distribution along
the extra dimension,
    to the black hole production cross-section
    for a $(4+1)$-dimensional model with all the standard model
particles localized at a single point,
    ($\sigma _{thick}/\sigma_{thin}$) as a function of the size of the extra dimension.
    }
    \label{cro4+1flat}
\end{figure}

Finally, we consider the thick brane case.
Strictly speaking, this is not one of the sub-classes of the split
fermion model.
Rather it is the case where
the brane on which the standard model fields are localized is not
infinitely thin
but has some finite thickness.
Here, we take  flat (homogeneous) distribution functions for
both quarks and the gluons.
Some characteristics are the same as for the previous two cases. When
the size of the
extra dimension is small compared to the size of the produced black
holes, there
is no difference between the thick and thin brane cases. As the
thickness of the brane increases, the cross section ratio decreases
very
quickly. But for large thickness of the brane
(much larger than typical black hole size), there is no specific size
of extra dimensions
(brane thickness) where the ratio stops decreasing sharply as in the
two previous cases.
 Instead, the cross section ratio decreases like $r_h/L$.
For higher $M_\star $, $r_h$ is smaller, and higer  $M_\star $ curves decay faster.
These features can be seen in Fig. \ref{cro4+1flat}.

%
%
\subsection{Angular momentum distribution}

\indent

The angular momentum of the black holes produced at the LHC cannot be
directly measured.
However, a black hole's angular momentum strongly affects its Hawking
radiation,
especially through the super-radiance effect described in the
introduction.
In most analysis of Hawking radiation from mini black holes,
the bulk component of angular momentum is neglected.
However, in the split fermion model, this assumption cannot be
justified.
Our results show that the bulk component of angular momentum can be as
large as the brane component,
thus greatly amplifying bulk radiation
(at least in the initial stages after formation, while the black hole
is still rotating fast). 

We must emphasize that the angular momentum analysis 
performed here is classical. It is likely
that the results and  even the conclusions will change 
if quantum corrections are not negligible.

\subsubsection{One extra dimension}

\begin{figure}[h!]
    \centering{
    \includegraphics[width=3.5in]{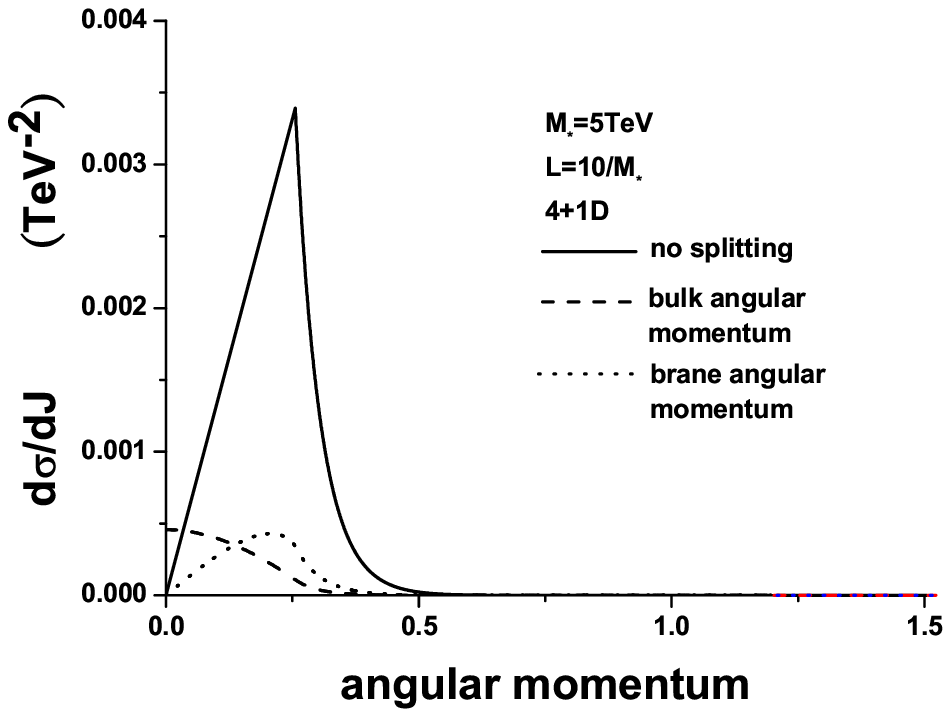} }
    \caption{
    Angular momentum distribution for a $(4+1)$-dimensional model.
    The fermions have Gaussian profile in the extra dimension. Gluons
have uniform (flat)
profile in extra dimensions. For comparison, we keep the plot in the
non-split case (solid line).
$M_{\star}=5TeV$, $L=10 M_{\star}^{-1}$.}
    \label{4+1JD}
\end{figure}

We first consider the $(4+1)$-dimensional case.
In Fig. \ref{4+1JD}, we plot the differential cross-section
$\frac{d\sigma}{dJ}$.
This encodes the expected number of black holes to be created as a
function of their angular momentum.
We take $M_{\star}=5TeV$ and fix the size of the extra dimension as
$L=10 M_{\star}^{-1}$. We we take the fermions to have Gaussian wave
 functions in the extra dimension.
The distributions of both bulk and brane components of the angular
momentum are plotted.
For comparison, on the same graph, we also show the distribution of the
brane component
of angular momentum for the the case where fermions are not split.

\begin{figure}[h!]
    \centering{
    \includegraphics[width=3.5in]{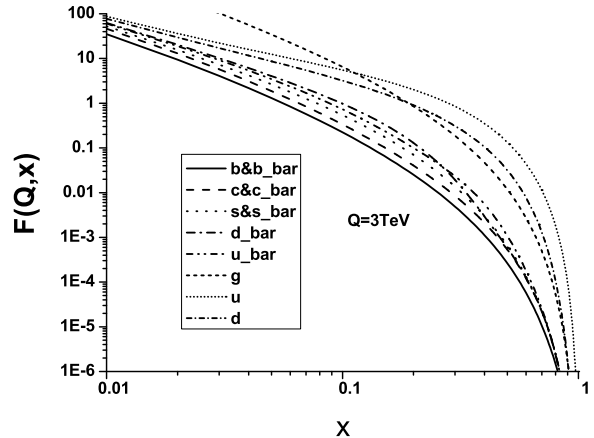} }
    \caption{Parton distribution functions}
    \label{pdfs}
\end{figure}

We see that the cross section in the non-split case increases with
angular momentum linearly.
After the angular momentum reaches the maximum that the smallest black
hole can provide
(most of the produced black holes will have the smallest possible
mass),
the cross section decreases very quickly.
To produce higher angular momentum black holes, one needs higher energy
partons.
From inspection of the parton distribution functions \cite{pdfs} (Fig.
\ref{pdfs}),
one can see that the number of higher energy partons decreases very
quickly.
Thus, at very high angular momenta, the cross section will be
suppressed.

As expected, in the split fermions case, the cross section  is reduced
compared to the un-split case.
The cross section still reaches its maximum when the angular momentum
in the brane directions
is comparable to the maximum that the smallest black hole can provide
(since most of the produced black holes will be the smallest possible).

Regarding the bulk component of angular momentum, since the cross
section will not be suppressed,
if the distance between partons in extra dimensions is zero,
zero angular momentum will have the highest cross section.
However, this is an artifact of one extra dimension.
This feature will change if there is more than one extra dimension.

\subsubsection{Two extra dimensions}

\begin{figure}[h!]
    \centering{
    \includegraphics[width=3.5in]{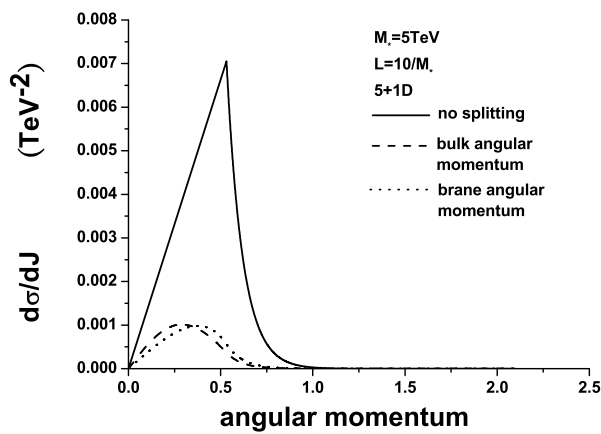} }
    \caption{
    Angular momentum distribution for a $(5+1)$-dimensional model.
    The fermions have Gaussian profile in the extra dimensions. Gluons
have uniform (flat)
profile in extra dimensions. For comparison, we keep the plot in the
non-split case (solid line).
$M_{\star}=5TeV$, $L=10 M_{\star}^{-1}$.}
    \label{5+1JD}
\end{figure}

We next consider the case with two extra dimensions.
The fermions again are taken to have Gaussian wave functions in the
extra dimensions.
>From figure \ref{5+1JD}, we see that the probability distribution of
the brane component of the angular momentum
behaves similarly as in $(4+1)$-dimensions.
Not so the bulk component of the angular momentum.
In particular, the most likely value of the bulk component of the
angular momentum is no longer zero.
When there are two extra dimensions, the probability to find two
particles at the same location
(which will yield  zero bulk angular momentum) becomes very small.
Therefore, though one obtains the highest cross section when particles
that are not separated in extra space collide,
this configuration is unlikely if there is more than one extra
dimension.

\begin{figure}
    \centering{
    \includegraphics[width=3.5in]{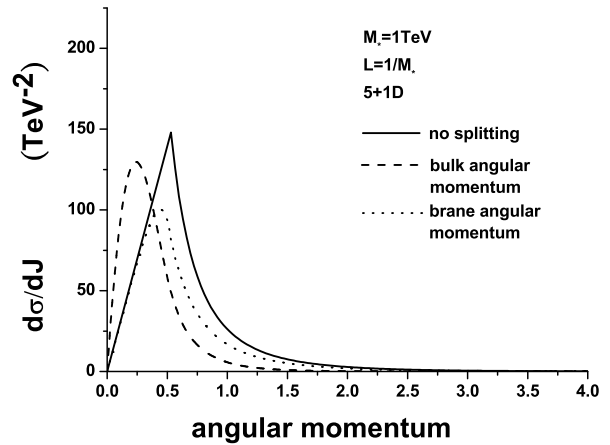} }
    \caption{Angular momentum distribution in 5+1D. The fermions and
gluons have uniform (flat)
profile in extra dimensions. For comparison, we keep the plot in the
non-split case (solid line).
    $M_{\star}=5TeV$, $L= M_{\star}^{-1}$.}
    \label{5+1JDf}
\end{figure}

Finally, in Fig. \ref{5+1JDf}, we show the case of a thick brane (flat
quark and gluon profiles).
The size of the extra dimensions is ${\cal O}(M_{\star}^{-1})$.
The brane and bulk angular momentum distributions are very similar in
this case.
We should first note that in any case most of the produced black holes
will have the smallest possible mass (production of heavier black holes
is suppressed
in proportion to their mass). At distances of the order of the
gravitational radius of
these small black holes (i.e. $\sim M_{\star}^{-1}$) one does not
distinguish much between the brane and bulk directions. Therefore, the
difference between
the bulk and brane angular momentum distribution is small in
this case.

%
%
%
\section{Conclusion}

\indent

Black hole production at the LHC may be one of the earliest signatures
of TeV-scale quantum gravity. This topic has been subject to an
extensive
investigation
\cite{HKM,1,2,3,4,5,6,7,9,10,11,12,13,14,15,16,17,18,19,20,
21,22,23,24,25,26,27,28,29,30,31,32,33,34,35,36,37,37.5,38,39,40,41,42,43,
44,45,46,47,48,49,50,51,52,53}.
In order to make serious predictions, one needs to
work within the context of realistic phenomenologically valid models.
In the first place
one has to suppress a very rapid quantum gravity mediated proton decay.
There are also many other problems related to TeV-scale quantum gravity
like large $n-{\bar n}$ oscillations, flavor changing neutral currents,
large mixing between leptons, {\it etc.}. It is widely accepted that
so-called
`split fermion" model (where fermions are localized at different points
in extra dimensions)
can solve most of these problems.

In this paper we re-examined black holes production and their angular
momentum
distribution in the context of this model with split fermions.
As a consequence of separation of partons in the extra dimensions, we
find that
the total production cross section is reduced compared with models
where all the fermions are localized in the same point in extra
dimensions.

Due to the non-zero impact parameter, most of the produced black holes
will be
rotating. Split fermions imply that the bulk component of the black
hole angular
momentum can not be neglected and must be taken into account in studies
of Hawking
radiation from such black holes.  In a follow on work (in preparation)
with collaborators
from the ATLAS consortium we will consider the impact of these and
related effects
on experimentally measurable quantities.

\vspace{12pt} {\bf Acknowledgements}:\ \

We thank Cigdem Issever, Nicholas Brett and Jeff Tseng of the Oxford
ATLAS group
for many productive conversations.  GDS is  supported in part by
fellowships
from the John Simon Guggenheim Memorial Foundation and the Beecroft
Institute
for Particle Astrophysics and Cosmology, Oxford.  He thanks the
Beecroft Institute
and Oxford Physics for their hospitality during the course of this
research.
This research is supported by a grant from the US DOE to the
particle-astrophysics
theory group at CWRU.

%
%

\end{document}